\documentclass{elsart}

\begin{document}
\def\v{$V_2 O_3$}
\def\n{$Ni {Se}_{x} S_{2-x}$}

\begin{frontmatter}
\title{{The Mott transition in $V_2 O_3$ and $Ni Se_x S_{2-x}$:} \\  Insights From Dynamical Mean Field Theory}
\author{G. ~Kotliar}

\address{ Serin Physics Laboratory \\ Rutgers
University, \\ Piscataway  New Jersey \\ 08855-0849, USA}
\begin{abstract}

We   discuss some aspects of 
the    pressure (or interaction)
driven 
Mott transition, in three dimensional transition metal oxides by means of dynamical
 mean field theory.
We isolate  
the universal  properties of the transition
from the aspects which depend more on the
detailed  chemistry
of the compounds.
In this light we can understand the main differences
and the remarkable similarities between the \n  and the \v 
system.
Both  theory  and experiment converge on 
the transfer
of spectral weight from low energies to high energies
as the universal mechanism underlying the Mott transition,
and we comment on the possible relevance of these ideas
to other metal to non metal transitions.

\end{abstract}

\end{frontmatter}

{\bf Introduction}
The Mott transition problem has fascinated a generation
of theorists, and  is believed to be realized in materials
straddling the localization delocalization boundary
such as
\def\materials{$V_2 O_3$ and $Ni Se_x S_{2-x}$}
\materials \cite{Wilson}.
Recently great theoretical progress  on this problem
has been achieved
using dynamical mean field methods\cite{Georges}.
In this paper we will discuss
how this approach gives insights into the physics of these
two materials. Recent detailed
experimental studies show that 
in spite of  the  many  similarities between these two materials,
some
physical quantities behave quite different as 
the metal to non metal
boundary is approached \cite{Takagi} \cite{Rosenbaum}.
Our goal  is to indicate  which aspects of the problem
can be  qualitatively understood from simple models that ignore the
detailed solid state chemistry of the compound (i.e. band structure
and orbital degeneracy),  and which aspects require a more realistic
description which takes into account those effects.
We  will then summarize the differences and similarities between
the two systems and how they can be  understood in a dynamical mean
field framework.
We conclude with a brief discussion of the applicability of these
concepts to other systems.

{\noindent \bf{Dynamical Mean Field  Theory in the  Continuum  }}
The dynamical mean field approach  (DMFT), in the continuum,
can be most easily explained by analogy with the well established
Density Functional Theory DFT.
In this approach,
$\rho(r)$, i.e. the density is the  basic quantity of the theory.
Then one defines an exact  functional {$\Gamma_{DFT}(\rho)$, of the density
such that its minimization  ${{\delta \Gamma_{DFT}}\over {\delta \rho
}}
(\rho_{phys})=0$ yields the physical density of the problem.
The reason why the DFT approach, sometimes fails in describing
Mott transitions, is that when these transitions involve very
small volume changes, as in the $NiS_2$ system, the density
near the transition does not change that much.
An approximate density functional, such as
$\Gamma_{LDA}$ then  fails to detect these changes.

Another way of rationalizing the failures of DFT in describing
Mott transitions, and other problems in strongly correlated electron
systems, given its extraordinary  success for weakly correlated materials,
is to notice that while DFT is in principle a theory of
ground properties, the excitation spectra of weakly correlated materials,
bears a great deal of similarity  to the  spectra of the Kohn Sham
equations, (band theory).
On the other hand, the excitation spectra of correlated electrons,
as measured in photoemission spectroscopy, has in addition to
quasiparticle-like, dispersing features, broader, atomic like, broader
and less  dispersive features
called
Hubbard bands.
There is no trace of the Hubbard bands in the Kohn Sham spectra.
A formalism suitable for correlated electron systems, should treat
both the quasiparticle features, and the Hubbard bands, {\it on the same
footing}. Formulating the problem in terms of the local spectral function
accomplishes precisely this goal.

The basic object in the DMFT approach is    the local Greens function, i.e.
the density of states for adding or removing one particle
which is measurable in photo-emission and inverse photo-emission spectroscopy.
One formulates the theory in terms of the local Greens Function ${G^L}$,
with 
r  and r' are  vectors in a Wigner Seitz cell.
\begin{equation}
 {G^L} (\tau, r, \tau',r')= -<T(\psi(r,\tau) {\psi^\dagger}(r',\tau')>
\end{equation}
Intuitively, we can think of DMFT as an attempt to
 "frequency resolve the density" since
$ \rho(r)= -{1 \over \pi} \int f(\omega) Im {{G^L}_R}(r,r, \omega)$.
In problems which are driven by transfer of spectral weight from one
frequency range to another, without appreciable changes in the total
density, the dynamical mean field method is likely to provide a more
adequate framework.

The construction of the dynamical mean field theory in the continuum
then proceeds by analogy with the density functional theory.
One proves the existence of a unique functional $\Gamma_{DMFT}[G]$ such that
${{\delta \Gamma_{DMFT} } \over {\delta G} }=0 $  at the physical local
Greens function G \cite{Chitra}.
Furthermore, this functional has the form,
\begin{equation}
\Gamma_{DMFT}[{G^L}]=\Gamma_{universal}[{G^L}]+ \int {v_c}(r) {G^L}(\tau=0^-,r,0^+,r)
\label{dmft}
\end{equation}
with the universal functional independent of the crystal potential,
${v_c}(r)$.
This is proved using Legendre transformation techniques
\cite{Chitra} and the existence proof is valid as long
as certain invertibility condition is satisfied.
Alternatively, one can formulate the problem in terms of  
a basic object ${G^L} (\tau, r, \tau',r')= -\sum_{n} {P } (R_n)<T(\psi(r+R_n,\tau) {\psi^\dagger}(r',\tau')>
$ related to the Greens function via  the 
localization kernel $P(R_n)\equiv {\int_{BZ}} dk   e^{i k . R_n} $.

From the point of view of  a practical  implementation,
we need explicit  forms to approximate  the  exact functional $\Gamma_{DMFT}$.
One possibility is to express the full  Hamiltonian in a complete
basis of Wannier functions built from the solution of the Hartree
Fock equations, truncating the dynamical part of the interaction leaving
only the local terms and applying to the latter the standard dynamical
mean field construction \cite{Georges}.
This procedure, is likely to be more accurate, the more localized the
Wannier basis is.
A  different construction of an approximate functional  which does not make
reference to an explicit basis  can be carried out using the cavity construction
directly in the continuum
\cite{Chitra}.
The study of model Hamiltonians using DMFT, is by now a vast subject \cite{Georges}.
The continuum  formulation goes beyond the treatment of model
Hamiltonians because   it allows  the determination
of the model Hamiltonian parameters in a self consistent fashion.
This formulation  necessary
to describe
the changes in the lattice parameters that accompany the metal to insulator
transition.

{\bf  Insights From Dynamical Mean Field Theory}
The DMFT, has systematized and unified seemingly different
approaches to the study  of correlated electron systems.
Furthermore it  has given rise to
several surprising  insights, that will guide us in the
interpretation of experimental  data \cite{Georges}.
The functional in eq. \ref{dmft} is highly non linear and in
certain region of parameters  has multiple minima.
The phase diagrams   
are constructed  by comparing the free energies of
these states.
In the presence
orbital degeneracy or other sources of magnetic frustration,
DMFT has  
several states (or dynamical  mean field solutions)
with {\it very different}
low energy  properties  which are
{\it very close} in energy.
It was found that in the presence of magnetic
frustration, the phase diagram of the  Hubbard 
Hamiltonian, has the same phases and topology
of that of \n and \v \cite{Thomas}.
Magnetic frustration 
suppresses magnetic long range order and 
allows  a first order paramagnetic metal to
paramagnetic insulator transition.
The  conjecture that the Hubbard model could produce
the  phase diagram   of \v  was put forward long ago \cite{Cyrot}, 
however     one lacked a  thermodynamically consistent treatment
of the model to demonstrate the point, and  it was not clear
that one needed a finite amount of  magnetic  frustration \cite{Wilson}.

The origin
of this magnetic frustration is very different in the two 
systems and  reflects the  different quantum chemistry
of  the two compounds.
$V_2 O_3$ is a 
a Mott Hubbard system while, $Ni Se_2$ is a charge transfer
but this is not the main source of differences
which lies in the different
kind of orbital degeneracy.
The minimal model for \v, due to Castellani et. al., involves 
one electron  in 
a doubly degenerate $a_{1g}$ orbital, so the Mott transition
has to lift, not only the spin degeneracy but also the
orbital degeneracy.
The \n system on the other hand is described by two holes 
in a twofold degenerate  $e_g$ orbital. 
Hunds rule coupling favors 
parallel spins and both orbitals occupied, 
 at each site lifting  completely 
the orbital degeneracy.
The source of frustration, in the \n case, is the fact that the Ni ions
in the pyrite structure
occupy  an fcc lattice which is  strongly frustrated  by the presence of  
nearest neighbor and   next nearest neighbor exchange constants of
similar order of magnitude.
The source of magnetic frustration in \v, is tied to the orbital degeneracy
, since  the exchange among orthogonal orbits, is ferromagnetic, while
the exchange among identical orbits is antiferromangetic, the magnetic
Hamiltonian in the ordered phase of \v has both ferro and anti-ferro
couplings, leading to magnetic frustration.

The finite temperature part of the phase diagram of \n and \v
is qualitatively similar, its physics is  captured by
a simple one band Hubbard model with frustration.
The low temperature phases, and the nature of the transition
into the low temperature phases, on the other hand, 
is different. In \n, the orbital
degeneracy is totally removed  and the spin
ordering transition (Paramagnetic Metal to Antiferromagnetic Insulator
or Paramagnetic Metal  to Antiferromagnetic Metal)  transition
is of the  second order.
In the \v system, the orbital degeneracy is lifted by
orbital ordering, which causes the exchange constants
to switch discontinuously \cite{Bao}.

The behavior in the paramagnetic phase, near the Mott transition, 
can be described in terms of a two fluid picture.
The spectral function of correlated electron
systems contains two kinds of features,
quasiparticles, and incoherent features.
They both play a crucial role at finite temperatures, near the Mott
transition.
When the Mott Transition is approached,
as a function of a control parameter (temperature,  pressure,
etc) 
spectral
weight  is transfered from low  energy  to high energy.
Several spectroscopic measurements in \n and \v are consistent
with these ideas. 
Optical measurements \cite{Thomas} in the metallic phase of \v
can be interpreted as a result of anomalous transfer of spectral
weight from low to high energies. 
Recent photoemission experiments by the group of Z. X. Shen \cite{Shen}
in the \n system 
have provided  direct evidence that  the   
temperature driven  metal to insulator transition point   
can   indeed be described as  
the evaporation of the quasiparticle peak, in the one particle
spectra.
Theoretically,  the anomalous temperature dependence  of the single particle
spectra was first  studied  in the paramagnetic
phase \cite{Georges}.  However  this seems to be a very robust feature of the strongly  correlated metallic regime near a  frustrated paramagnetic insulator
and  also occurs  in  antiferromagnetic metallic phases \cite{Watanabe}
\cite{Chitra1}.
The transfer of spectral weight as the mechanism driving the Metal to
Insulator transition, is quite different  from the  opening of Slater
gap, which conserves the k integrated low energy spectral weight.
Finally it is worth mentioning that a comparison between the photo emission
spectra of the paramagnetic metal and the anti ferromagnetic insulating
phase of \v, is also consistent with the transfer of spectral weight
from low energies to extremely high energies \cite{Allen}.


{\bf Mass Enhancement}   
In \v the mass increases as a function of pressure until the
metal to insulator transition is reached.
On the other hand in \n the mass increases as pressure is applied,
and starts decreasing after it entered the metallic anti-ferromagnetic
phase \cite{Takagi} \cite{Rosenbaum}. To understand this contrasting behavior,  it is useful
to  start from 
general Fermi liquid considerations, 
in a metallic
phase with magnetic long range order.
The one particle Greens function, is separated into a singular part
and a smooth incoherent part.
\begin{equation}
G_{\sigma \alpha} (\omega, k) \approx { {Z_{k \sigma \alpha }} \over {\omega- E_{k
\sigma \alpha }}} + G_{inc, \sigma \alpha} (\omega, k)
\end{equation}

The Fermi Surface, is defined by the locus of zeros of the quasiparticle
energy, i,e.
 $ E_{k \sigma \alpha }=0 $ and has multiple sheets in  orbitally degenerate
systems, $\sigma$ and $\alpha$ are spin and orbital indices.
The linear term of the specific heat is proportional to the
quasiparticle density of states  
\begin{equation}
\gamma \propto
{\sum_{\sigma, k} \delta (E_{\sigma, k \alpha})}= \sum_{\sigma} A_{\sigma \alpha} <
{m^*}_{\sigma, k \alpha}>
\end{equation}
Here we have defined an effective mass by,
${{\partial E_{k \sigma \alpha}} \over {\partial {k_{||}}}}= {{k_F
}(\theta)  \over{{m_{\sigma \alpha}}^*(\theta)}}$,  
$A(x)_{\sigma \alpha}$ denotes the  area of the Fermi surface parametrized
by the angle $\theta$, and
$< >$ denotes an average over the Fermi surface.

The basic physics controlling the behavior of the  specific heat 
near  the AM to AI phase boundary 
is the 
competition between  the
 the increase of
$<  {m_{k\sigma \alpha}}^*> $   
and the decrease of
the Fermi surface area $A  $ .
Quantitative calculations of the specific heat displaying these
tendencies 
have been performed in a frustrated version of the one band
Hubbard model
\cite{Chitra1}.
In the  anti-ferromagnetic  phase, 
$ m^*  \propto  [1- {{\partial \Sigma(i \omega)}
\over {\partial i \omega}}] = Z^{-1}$
and its increase
is  cut off by staggered magnetization \cite{Chitra1}
On the other hand, 
$A $ decreases due to self consistent change in band structure.
When the  staggered magnetization is
large  the change in area is the dominant effect,
whereas the first effect
dominates when the magnetism is weak.

The different behavior of the mass enhancement as the AM 
(antiferromagnetic metal) to AI (anti-ferromagnetic insulator)  boundary
is approached can be understood if we assume that the   
metallic magnetism is stronger in  \n  than in
$V_2 O_3$.
The first effect  dominates in $V_2 O_3$  while 
the second effect 
dominates in   $Ni (S Se)_2$ 
as a result 
$\gamma$ increases (decreases)  
as the MIT is approached  
 in the two materials respectively.
This is consistent with the larger size of the  magnetic moment
of the \n compared with the \v system as 
observed in neutron scattering \cite{Miyadai}. 
This is also 
consistent with  the DMFT interpretation
of the  low temperature Hall coefficient measurements which are 
given by the  Boltzman formula,   using  $ E_{k \sigma \alpha}$
as the quasiparticle energy. 
In \v the low temperature Hall coefficient  depends very weakly
on pressure, while in \n the Hall coefficient increases dramatically
as the pressure driven MIT is approached.

{\bf Conclusion}
The main shortcoming of the dynamical mean field approximation is
the neglect of  short range magnetic correlations in the paramagnetic phase.
It is most reliable  when  the  effective magnetic interactions are
frustrated or  small.
The local 
approximation might  also  be an  appropriate starting point to describe 
the recently discover  metal to insulator transition in
two dimensional mosfets  \cite{Kravchenko}, in the accessible temperature
range.  
In the low density regime of the electron gas and
in the presence of disorder,
the exchange constants
in the insulating phase are very small and  frustrated.
It is interesting to notice  that  the conductivity  data  
in Mosfets \cite{Kravchenko}
and in the \n system \cite{Takagi} are remarkably similar, once we recognize
that in the \n  system the bare kinetic energy is a two orders of magnitude
larger.


\begin{thebibliography}{20}
\bibitem{Wilson}J.  Wilson in the metallic and nonmetallic states of matter.
Edited by P.P. Edwards and C. N.R. Rao, 215 (1985). 
\bibitem{Georges} for a recent review  and additional references see
A. Georges et. al. Reviews of Modern  Physics 68, (1996) 13. 
\bibitem{Takagi} S. Miyasaka et. al submitted to PRL.
\bibitem{Rosenbaum} S. Carter et. al. Phys. Rev. Lett 67, 3440 (1991).
\bibitem{Chitra} R. Chitra and G. Kotliar ( preprint).
\bibitem{Chitra1} R. Chitra and G. Kotliar Phys. Rev. Lett 83, 2386 (1999).
\bibitem{Bao}W. Bao et. al. Phys. Rev. Lett. 78, (1997) 507. 
\bibitem{Shen} Matsuura et. al. Phys. Rev. B  58, 1998, 3690.
\bibitem{Cyrot}
M. Cyrot  and P. Lacour-Gayet Solid State Comm, 11 (1972) 1767.
\bibitem{Watanabe} H. Watanabe and  S. Doniach
Phys. Rev. B 57, 3289 (1998).
\bibitem{Thomas} M. Rozenberg et. al. Phys. Rev. Lett 75, 105 (1995)
\bibitem{Allen} J. Allen in  The Hubbard Model
 edited by D. Baeriswyl, D.K. Campbell, J.M.P. Carmelo, F. Guinea
and E. Louis (Plenum, New York, 1995), 357. 
\bibitem{Miyadai} S. Sudo et. al. Jour. Mag. and Mag. Mat. 114, (1992), 57. 
\bibitem{Kravchenko} S. Kravchenko et. al. PRB 51, 7038 (1995). 


\end{thebibliography}
\end{document}